\documentclass[runningheads,a4paper]{llncs}

\usepackage{amssymb}
\usepackage{amsmath}
\usepackage{graphicx}
\usepackage{multirow}
\usepackage{algorithm2e}
\usepackage{url}

\begin{document}

\mainmatter  

\title{Resilience of Social Networks Under Different Attack Strategies}

\titlerunning{Resilience of Social Networks}

%
%
\author{Mohammad Ayub Latif\inst{1} \and Muhammad Naveed\inst{2} \and Faraz Zaidi\inst{1,3}}

\authorrunning{Latif et al.}

\institute{Karachi Institute of Economics and Technology, Karachi, Pakistan.\\ 
		   \email{malatif@pafkiet.edu.pk, faraz@pafkiet.edu.pk}
\and		
			Muhammad Ali Jinnah University, Karachi, Pakistan\\
			\email{naveed.shaikh@jinnah.edu}
\and			
			University of Lausanne,	Lausanne, Switzerland			
			}

%
%

\toctitle{Lecture Notes in Computer Science}
\tocauthor{Authors' Instructions}
\maketitle

\begin{abstract}

Recent years have seen the world become a closely connected society with the emergence of different types of social networks. Online social networks have provided a way to bridge long distances and establish numerous communication channels which were not possible earlier. These networks exhibit interesting behavior under intentional attacks and random failures where different structural properties influence the resilience in different ways.

In this paper, we perform two sets of experiments and draw conclusions from the results pertaining to the resilience of social networks. The first experiment performs a comparative analysis of four different classes of networks namely small world networks, scale free networks, small world-scale free networks and random networks with four semantically different social networks under different attack strategies. The second experiment compares the resilience of these semantically different social networks under different attack strategies. Empirical analysis reveals interesting behavior of different classes of networks with different attack strategies.

\end{abstract}

\section{Introduction}\label{sec::introduction}

Online communication channels or mediums of computer mediated communication such as emails, blogs and online social networking websites represent different forms of social networks. These networks have attracted billions of users in recent years \cite{ahn07} adding new dimensions to socializing behavior and communication technologies. These networks provide a challenging opportunity for researchers from different domains to analyze and understand how the new age of communication is shaping the future. These networks also help us understand how information disseminates \cite{iribarren11} in social networks and how communication plays a role in the creation of new knowledge \cite{prusak01}.

An important aspect of these networks is that they can undergo intentional attacks or random failures which results in communication breakdown. Thus researchers have focussed on studying the stability of networks in terms of how resilient or how robust these networks are against any malicious activity or natural random failures \cite{cohen00,holme02a}. Given a network with $n$ nodes and $m$ edges, targeted or random attacks are modeled by the removal of a series of selected nodes or edges from the network. The way these nodes or links are chosen, known as the attack strategy determines the impact and behavoir it causes on the resilience of the network.

The natural evolution of these networks has introduced several structural properties which play an important role in determining the resilience of these networks. These properties characterize the behvior of many social and other real world networks giving us two important classifications, the scale free networks \cite{barabasi99} and small world networks \cite{watts98}. Scale free networks have a degree distribution following power law\footnote{power law $p_k \sim k^{-\alpha}$ where $\alpha$ is usually in the range of $[2,3]$}. Small world networks have low average path lengths (APL) scaling logarithmically with the increase in number of nodes (n) and high clustering coefficient implying the presence of large number of triads present in the network. Many social networks have both these structural properties giving us another class of networks called, small world-scale free networks.

Scale free networks have been extensively studied with respect to resilience, and Internet provides the perfect dataset for such analysis \cite{cohen00,cohen01,guillaume05}. Researchers have shown that scale free networks are highly sensitive to targeted attacks and very robust against random attack strategies \cite{cohen00,cohen01}. This phenomena is often termed as the `Achilles heel of the Internet'. Resilience of networks with only small world properties, and both small world-scale free properties has not been the focus of studies even though many social networks around us exhibit both small world and scale free properties \cite{albert02,newman06a}. 

One example of such networks is the structure of the world wide web studied by \cite{broder00}. The authors found that the web has a bow tie structure and is very robust against targeted attack on nodes. This result contradicts the findings that scale free networks are fragile to targeted attacks. The reason is that deleting nodes with high degree is not enough to cross the percolation threshold as the average edge-node ratio (also called density or average degree) of these graphs is very high. This finding is similar to our results for the case of social networks.

In this paper, we perform two sets of experiments. The first set of experiments compares the behavior of four different classes of networks, small world networks, scale free networks, small world-scale free networks and random networks with four equivalent size real social networks. These social networks are from a political blog, Epinions who-trust-whom network, Twitter social network and Co-authorship network of researchers. We study these networks under six different attack strategies which are, targeted attack on nodes and edges, random failure of nodes and edges, and almost random failure\footnote{defined in section \ref{sec::experimentation}} of nodes and edges \cite{guillaume05}. The idea is to see how structural organization of these different networks impact resilience when their edge-node ratio is equivalent to that of semantically different social networks. Our results lead us to these findings:

\begin{itemize}
\item Five of the six attack strategies behave similarly for all different classes of networks, the exception being targeted attack on nodes.
\item Clustering coefficient has no effect on the resilience of networks if netoworks with high edge-node ratio are studied.
\item Results show scale free and small world-scale free networks are more fragile to targeted attacks. Targeted attack on edges removes the same number of edges from other classes of networks and the behavior of all classes including random networks remains the same indicating that the different behavior in scale free and small world-scale free networks is due to the large number of edges being removed from the network and not due to the structural organization of the network itself.
\item Network generation models used to generate small world, scale free, and small world-scale free networks differ largely from the behavior of real networks in terms of resilience suggesting structural flaws in existing network generation models.
\end{itemize}

The second experiment studies the resilience of the four real social networks in terms of different attack strategies on nodes,  which was found to be more interesting in the previous experiment. The results can be summarized below:

\begin{itemize}
\item We observe only minor differences between random and almost random failures for blog, epinions and twitter networks as compared to the author network which demonstrates some differences between the two strategies.
\item Attack on Targeted nodes clearly differs from random and almost random failures whereas the author network seems to be the most fragile. The blog, epinions and twitter network demonstrate graceful degradation in performance in terms of size of biggest component.
\end{itemize}

Rest of the paper is structured as follows: In the next section we discuss several studies pertaining to resilience of different types of networks. Section \ref{sec::data} provides the details of the real world datasets and the networks generated using different network generation models. In section \ref{sec::experimentation}, we describe our experimental set up and the metrics used for analysis. Section \ref{sec::results} explains the results obtained and provide findings from the experimentation and finally we conclude in section \ref{sec::conclusion} also giving future research directions.

\section{Related Work}\label{sec::related}

One of the earliest studies to demonstrate that scale free networks are more robust against random failures was conducted by \cite{albert00}. The authors also discuss the vulnerability of scale free networks to targeted attacks. Cohen \textit{et al.} \cite{cohen00,cohen01} study the resilience of internet under random and targeted attacks on nodes. For the case of random attacks, they conclude that even after $100\%$ removal of nodes, the connectivity of the biggest component remains intact that spans the whole of the network. The authors claim that this condition will remain true for other networks if their connectivity distribution follows power law with power law coefficient less than 3. For the case of targeted attacks, scale free networks are highly sensitive to targeted attacks on nodes as the biggest connected component disintegrates much sooner. Holme \textit{et al.} \cite{holme02a} study attacks on edges using betweenness centrality where edges with the highest centrality are removed. They show that recalculating betweenness centrality after each deletion is a more effective attack strategy for complex networks. Paul \textit{et al.} \cite{paul04} discuss that networks with a given degree distribution may be very resilient to one type of failure or attack but not to another. They determine network design strategies to maximize the robustness of networks to both intentional attacks and random failures keeping the cost of the network constant where cost is measured in terms of network connections. Analytical solutions for site percolation on random graphs with general degree distributions were studied by \cite{callaway00} for a variety of cases such as site and bond percolation. Serrano \textit{et al.} \cite{serrano06b} introduce a framework to analyze percolation properties of random clustered networks and small world-scale free networks. They find that the high number of triads can affect some properties such as the size and resilience of biggest connected component. Wang \textit{et al.} \cite{wang06} studied the robustness of scale free networks to random failures from the perspective of network heterogeneity. They examine the relationship of entropy of the degree distribution, minimal connectivity and scaling component obtaining optimal design for scale free networks against random failure. Estrada \cite{estrada06} studied sparse complex networks having high connectivity known as good expansion. Using a graph spectral method, the author introduces a new parameter to measure the good expansion and classify 51 real-world complex networks into four groups with different resilience against targeted node attacks. Wang and Rong\cite{wang09} analyse the response of scale free networks to different types of attacks on edges during cascading propagation. They used the scale free model \cite{barabasi99} and reported that scale free networks are more fragile to attacks on the edges with the lowest loads than the ones with the highest loads. Liu \textit{et al.} also affirm  that scale free networks are highly resilient to random failures. The authors suggest network design guidelines which maximize the network robustness to random and targeted attacks. A comprehensive study conducted by Magnien \textit{et al.}\cite{magnien11} survey the impact of failures and attacks on Poisson and power law random networks considering the main results of the field acquired. The authors also list new findings which are stated as under:
\begin{itemize}
\item Focusing on the random failure of nodes and edges, although previous researchers had predicted completely different behavior for Poisson and power law networks, in practice the differences, are vital but not huge. Our results re-enforce these results specially for the case of social networks.
\item  The authors also invalidate the explanation that targeted attacks are very efficient on power-law networks because they remove many links, random removal of as many links also result in breakdown of the network.
\item Networks with Poisson degree distribution behave similarly in case of random node failures and targeted attacks, it must be noted that their threshold is significantly lower in the second case. This goes against the often claimed assumption that, because all nodes have almost the same degree in a Poisson network, there is little difference between random node failures and targeted attacks. 
\end{itemize}

Resilience has not been extensively studied for social networks. Moreover, studies focus on networks that are either only scale free or their sizes are not comparable to online social networks readily available around us. Considering the new findings that deviate with the previous results, we get a strong motivation to further investigate resilience of different types of complex networks with a focus on social networks. Our empirical results reaffirm most of these findings of \cite{magnien11} where our focus is on semantically different social networks.

\section{Data Sets}\label{sec::data}

We have used four semantically different real world networks which represent social communication of different forms. These are the Political Blog network, Twitter, Epinions and Author network which we are abbreviated as (RN) and are described below.

\textit{Political Blog} network is a network of hyperlinks between weblogs on US politics, recorded in 2005 by Adamic and Glance\cite{adamic05}. \textit{Twitter} network is one of the most popular online social networks for communication among online users and we have used the dataset extracted by \cite{hashmi12}. \textit{Epinions} network is a who-trust-whom online network of a customer analysis website Epinions and the data is downloaded from the stanford website ({\url{http://snap.stanford.edu/data/}}) where it is publicly available. The \textit{Author} network is a co-authorship network where two authors are linked with an edge, if they co-authored a common work(an article, book etc). The dataset is made available by Vladimir Batagelj and Andrej Mrvar: Pajek datasets (\url{http://vlado.fmf.uni-lj.si/pub/networks/data/}). For all these networks, we only consider the biggest connected component and treat these networks as simple and undirected. Table \ref{tbl::networks} shows the number of nodes and edges in these networks along with the edge-node ratio. For each of these real  networks, we generated equivalent size networks using four network generation models referred above. The introduction of real data not only allowed us to select realistic edge-node ratio, but also to compare these models with real data. 

We have also used four network generation models to represent different types of networks. The small world (SW) model of Watts and Strogatz\cite{watts98}, the scale free (SF) model of Barabasi and Albert\cite{barabasi99}, the Small world-Scale free (HK) model of Holme and Kim\cite{holme02} and, the Erd\"{o}s (RD) model for Random graphs. 

The small world model can be tuned to the desired number of nodes and edges by initializing a regular graph where each node has a degree of $n$. The scale free model can be tuned by the number of edges each new node has in the network where all nodes connect preferentially. Similarly the model for small world-scale free networks can be tuned by the number of nodes each new node connects to, giving us a network with the desired edge-node ratio approximately. A random network is generated using n nodes and m edges where the degree distribution $p_k$ of the network follows a Poisson distribution $p_k = e^{-\lambda} \frac{\lambda^k}{k!}$. The networks we generated all had $\lambda > 1$ which signifies that most nodes in the network have a degree close to the mean degree of the network.

\begin{table}
\centering
\begin{tabular}{|l|c|c|c|}
\hline 
Network & Nodes & Edges & Edge-Node Ratio \\
\hline \hline
Blog & 1222 & 16714  & 13.6 \\
\hline
Twitter & 2492 & 17658  & 7.0  \\
\hline
Epinions & 2000 & 48720 & 24.3 \\
\hline
Author & 3621 & 9461 & 2.6 \\
\hline
\end{tabular}
\caption{Network Statistics for different social networks} \label{tbl::networks}
\end{table}

For the purpose of experimentation and empirical analysis, we generated 5 artificial networks each for small world, scale free, small world-scale free and random networks equivalent to the 4 different social networks giving us a total of 80 networks. We averaged the readings obtained for these networks although the networks had very little variations with standard deviations of less than 1 in all the cases. Table \ref{tbl::metrics} shows the degree of most connected nodes, clustering coefficients and average path lengths for the generated networks in comparison to real networks.

A clear similarity among all these networks is the low average path length which indicates that on average, nodes in all these networks lie close to each other following the famous `six degrees of separation' rule. All the real networks are both small world and scale free in nature, the scale free networks have a low clustering coefficient and the degree distribution of small world networks and random networks follow a Poisson distribution with $\lambda >1$. 

\begin{table}
\centering
\begin{tabular}{|l|c|c|c|c|c|}
\hline 
Data Set & {\ Real Network \ } & {\ RD \ } & {\ SW \ } & {\ SF \ } & {\ HK \ } \\
\hline \hline
& \multicolumn{5}{|c|}{Highest Degree of a Node\ } \\
\hline \hline 
Blog & 351 & 46 & 47 & 211 & 321 \\
\hline
Twitter & 237 & 27 & 27 & 253 & 319 \\
\hline
Epinions & 1192 & 77 & 72 & 373 & 560\\
\hline
Author & 102 & 15 & 16 & 201 & 183 \\
\hline
\hline

& \multicolumn{5}{|c|}{Clustering Coefficient\ } \\
\hline \hline
Blog & 0.32 & 0.02 & 0.56 & 0.07 & 0.24 \\
\hline
Twitter & 0.13 & 0.005 & 0.49 & 0.03 & 0.27 \\
\hline
Epinions & 0.27 & 0.02 & 0.58 & 0.08 & 0.22\\
\hline
Author & 0.53 & 0.001 & 0.31 & 0.01  & 0.42 \\
\hline
\hline

& \multicolumn{5}{|c|}{Average Path Length\ } \\
\hline \hline
Blog & 2.7 & 2.5 & 3.2 & 2.4 & 2.2 \\
\hline
Twitter & 3.4 & 3.2 & 4.2 & 2.9 & 2.8 \\
\hline 
Epinions & 2.2 & 2.2 & 3.0 & 2.2 & 2.0\\
\hline
Author & 5.31 & 5.07 & 6.41 & 3.4 & 4.0 \\
\hline
\multicolumn{6}{|c|}{\ } \\
\hline

\end{tabular}
\caption{Rd=Random Network, Sw=Small World, Sf=Scale Free, Hk=Holme and Kim Model for small world-scale free networks. Table shows different metrics calculated for the real and artificially generated networks for comparison.} \label{tbl::metrics}
\end{table}

\section{Experimentation}\label{sec::experimentation}

As described above, we studied resilience considering six attack strategies, three of which are for nodes and three for edges. These are Targeted attack on Nodes, Random failure of Nodes, Almost Random failure of Nodes, Targeted attack on Edges, Random failure of Edges and Almost Random failure of Edges. Each of these strategies is described below:

\textbf{Targeted attacks on nodes and edges:}  The attack strategy for targeted removal of nodes removes nodes in decreasing order of their degree (connectivity). This strategy is used by many other researchers\cite{guillaume05} for such studies.

To determine targeted edges, we propose a slightly different version from the one used by \cite{guillaume05}. The authors removed edges connected to high degree nodes which suits well for networks like scale free networks. Our method is inspired by the concept of funneling in social networks \cite{newman01b} where most connections of a person to other people are usually through a small set of people and connections with one or two famous personalities reduces the distance from all other people in the social network. Thus important edges linking many people would be the ones between high degree people. We assign a weight $\mathit{W}(e_{i,j})$ to all $m$ edges where $i,j$ represents the edge between nodes $i$ and $j$ based on the degree of each node using the  equation: $ \mathit{W}(e_{i,j})= deg(i) + deg(j)$. Nodes are removed in decreasing order of $\mathit{W}$ in an attempt to remove edges that connect most connected people in the network.

\textbf{Random failure of nodes and edges:} Random removal of nodes and edges is modeled by a series of failures of nodes or edges selected randomly from the network with equal probability. 

\textbf{Almost random failure of nodes and edges:} These attack strategies were described by \cite{guillaume05} as more efficient attack strategies in case of scale free networks. Almost random failure of nodes removes randomly selected nodes with degree atleast 2 and almost random failure of edges removes edges between vertices where the degree of each vertex is atleast 2.


\textbf{Quantifying Resilience of a network:} In order to quantify the resilience of a network, we use the two most commonly applied methods, one measures the number of nodes and the other measures the average path length of the biggest connected component in the network after each attack . The percentage of nodes still connected after an attack provides an estimation of how resilient networks are. Similarly the increase in the average distance from any one node to the other also provides an estimation of how resilient the networks are after each attack. We have studied the effects after every $10\%$ removal of either nodes or edges against the percentage of nodes remaining in the biggest connected component and the average path length of this component.

\section{Results and Discussion}\label{sec::results}

\begin{figure}
\begin{center}
\includegraphics[width=0.99\textwidth]{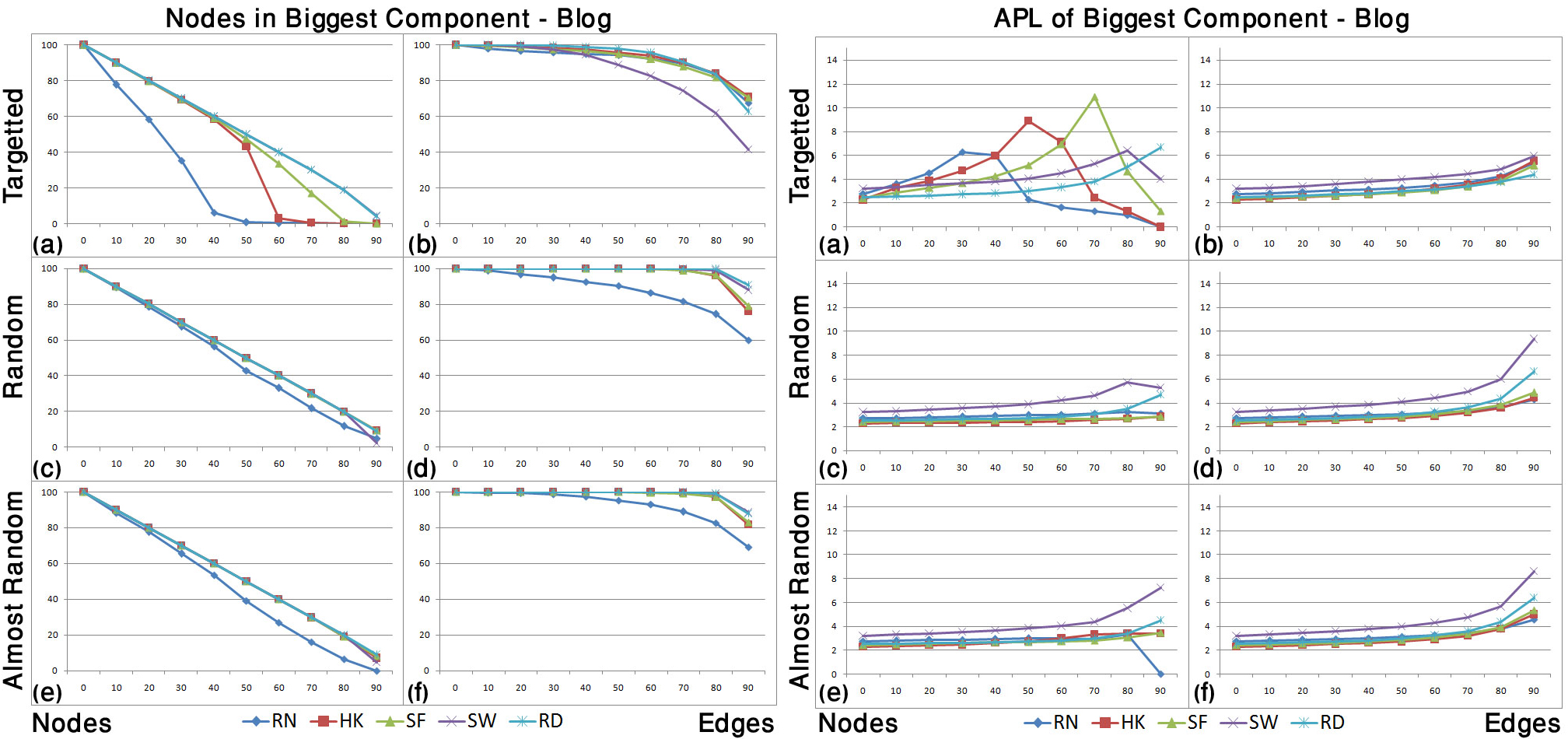}
\end{center}
\vspace{-20pt}
\caption{RN=Blog, HK=Small world-scale free, RD=Random, SF=Scale free, SW=Small world. X-axis: $\%$ of nodes (a,c,e) and edges (b,d,f) removed from the network,Y-axis: $\%$ of nodes (left) and APL (right) of the biggest connected component.}
\label{fig::blog}
\end{figure}

\begin{figure}
\begin{center}
\includegraphics[width=0.99\textwidth]{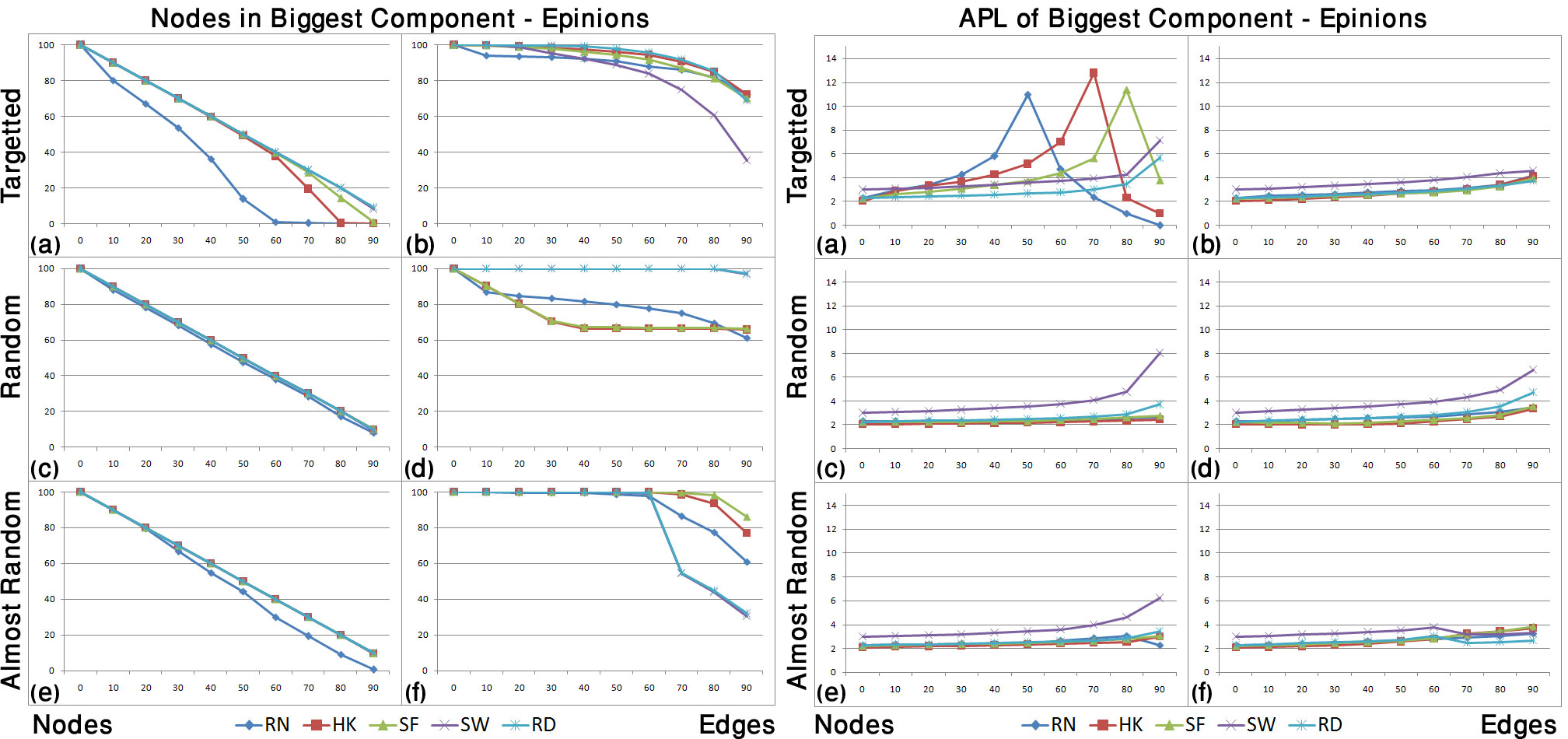}
\end{center}
\vspace{-20pt}
\caption{RN=Epinions, HK=Small world-scale free, RD=Random, SF=Scale free, SW=Small world. X-axis: $\%$ of nodes (a,c,e) and edges (b,d,f) removed from the network, Y-axis: $\%$ of nodes (left) and APL (right) of the biggest connected component.}
\label{fig::epinions}
\end{figure}

\begin{figure}
\begin{center}
\includegraphics[width=0.99\textwidth]{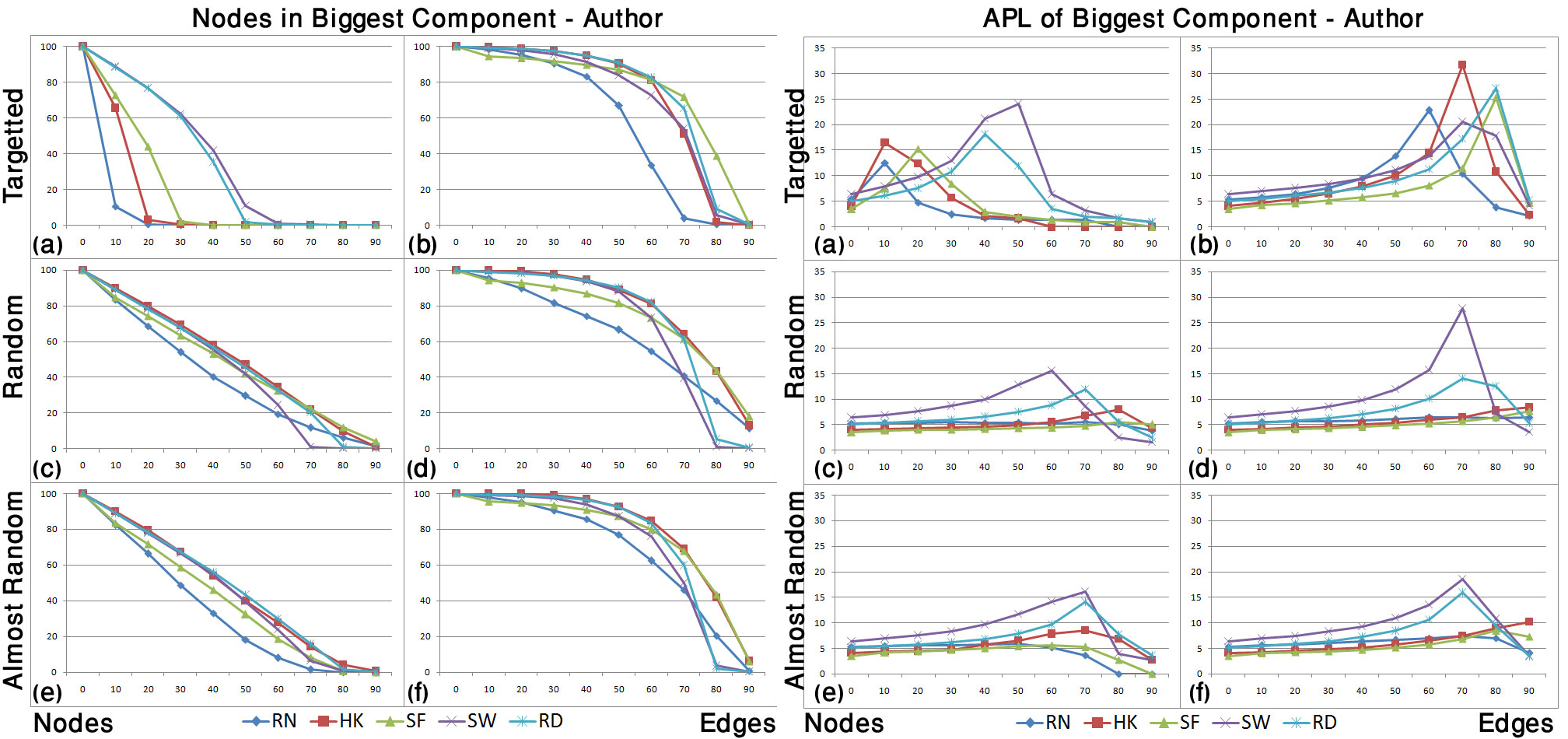}
\end{center}
\vspace{-20pt}
\caption{RN=Author, HK=Small world-scale free, RD=Random, SF=Scale free, SW=Small world. X-axis: $\%$ of nodes (a,c,e) and edges (b,d,f) removed from the network, Y-axis: $\%$ of nodes (left) and APL (right) of the biggest connected component.}
\label{fig::author}
\end{figure}

\begin{figure}
\begin{center}
\includegraphics[width=0.99\textwidth]{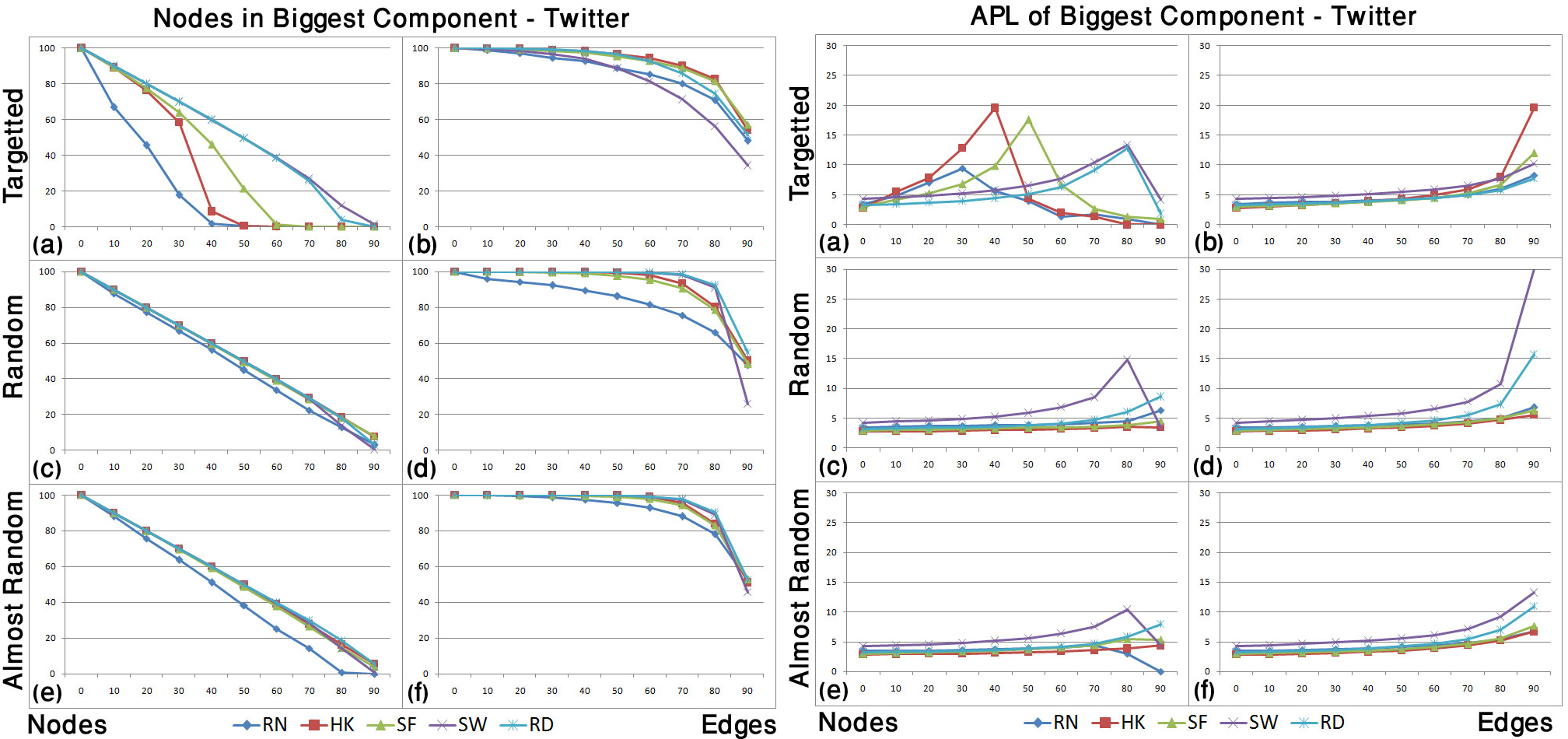}
\end{center}
\vspace{-20pt}
\caption{RN=Twitter, HK=Small world-scale free, RD=Random, SF=Scale free, SW=Small world. X-axis: $\%$ of nodes (a,c,e) and edges (b,d,f) removed from the network, Y-axis: $\%$ of nodes (left) and APL (right) of the biggest connected component.}
\label{fig::twitter}
\end{figure}

Figures (\ref{fig::blog}, \ref{fig::epinions}, \ref{fig::author} and \ref{fig::twitter}) show the results for all the four datasets with different attack strategies on nodes and edges. The first findings are for the cases where we studied targeted attack on edges, random attacks on nodes, random attacks on edges, almost random attack on nodes and almost random attack on edges. For all the these networks, we find that the real networks behave similarly to all 4 classes of networks, small world, scale free, small world-scale free and random networks if the same fraction of nodes or edges are removed as shown in figures. We justify these results based on the idea that increasing the minimum mean connectivity of nodes increases the robustness of networks to targeted and random attacks also discussed by \cite{liu05a}. For all the social networks under consideration, they have very high average connectivity as shown in Table \ref{tbl::networks}. Even for the case of author network which has an average connectivity of $2.6$, it is still high as compared to internet networks previously studied in the literature. 

Another generic finding is with respect to the clustering coefficients of different networks. Although there are extreme differences in random and social networks, having low values of even $0.001$ and high values of around $0.5$ (see Table \ref{tbl::metrics}) respectively. Still the behavior in terms of resilience remains the same for all these networks. This indicates that the presence or absence of triads does not reflect on the robustness of a network.

The analysis of scale free and small world-scale free networks which are fragile to targeted attacks when compared to small world and random networks is also very interesting. This result is the direct implication of the large number of edges removed from scale free and small world-scale free networks as a result of targeted attack on high degree nodes. Since nodes with very high degree are absent from small world and random networks, the same fraction of edges is not removed upon removal of high degree nodes and they give an impression that they are more resilient to targeted attack on nodes. The experiment on targeted attack on edges provides a contradiction to this result as equal number of edges are removed from real networks, scale free, small world, small world-scale free and random networks and the results show that the behavior of all these networks is almost the same. This claim is further justified from our results of random attack on edges as, again, they reveal similar behvior for all these classes of networks both in terms of size of biggest connected component and APL.

Another important result is the behavior of network generation models against the real networks for targeted attack on nodes. All the real world networks disintegrate more quickly than the artificially generated networks for all the four data sets used for experimentation. This observation highlights the fact that network generation models fail to accurately capture all the structural properties of real world networks. This is due to the structural organization of real social networks as compared to the artificially generated networks. In real networks, there is a high percentage of low degree nodes that are connected through high degree nodes only, when these high degree nodes are removed in case of targeted attacks, they immediately become disconnected. On the other hand, artificially generated networks are all based on random connectivity among nodes, and they are not necassarily connected only through high degree nodes, which makes them more resilient when high degree nodes are removed from the network.







We discuss the results for each set of first experiment below:

\textbf{Targeted attacks on Nodes:} As a general trend, both random and small world networks behave almost similarly for all the datasets. Further more, they are more resilient than small world-scale free (HK) networks and scale free networks (SF). Another important discovery is the behavior of all the real data sets in comparison to the artificially generated networks. Real datasets disintegrate faster than any other model as shown in figures \ref{fig::blog}(left), \ref{fig::epinions}(left), \ref{fig::author}(left) and \ref{fig::twitter}(left) for the case of targeted attacks. The least resilient network is the Author network which disintegrates after $10\%$ highest degree nodes are removed. For the case of Blog and Twitter network, around $40\%$ removal of high degree nodes is sufficient to break the entire network as the size of the biggest component falls below $10\%$, whereas epinions network falls below $10\%$ after around $50\%$ removal of nodes making it more resilient to targeted attacks. Again the edge-node ratio plays an important role as clearly epinions network has the highest value of $24.3$. This, when compared with the Author network with edge-node ration of $2.6$ indicates how having more edges nullifies the effects of targeted attacks on networks. The above similarity in the behavoir of networks is further reinforced after looking the behavoir of APL in figures \ref{fig::blog}(right), \ref{fig::epinions}(right), \ref{fig::author}(right) and \ref{fig::twitter}(right) where variations can only be observed in the case of targetted attack on nodes.

\textbf{Random failure of Nodes:} The behavior of random removal of nodes for all the six cases reveals an interesting similarity specially for the case of generated scale free network, small world network, random network and the real data sets. Particularly for the Blog and Epinions data, almost $100\%$ similar behavoir is evident from figure \ref{fig::blog}(left) and figure \ref{fig::epinions}(left). Twitter and Author networks also show high similarity as shown in figures \ref{fig::author}(left) and \ref{fig::twitter}(left). A linear decay is observed in the number of nodes present in the biggest component against linear removal of nodes which suggests that the nodes remain connected even after $90\%$ of the nodes are removed which demonstrates very high resilience for all these networks againt random node failures. The APL of small world networks for all data sets has a slightly higher value indicating minor difference in the empirical values, but the overall behavior and decay pattern is the same for all networks. 

\textbf{Almost Random failure of Nodes:} Just as random failures, almost random failure of nodes demonstrates a high similarity among the different classes of networks and the real networks. Differences can be observed only for the case of author network which has a much lower edge-node ratio. The behavior of the real author network deviates slightly from the other classes of networks. This is contradictory to the results of \cite{guillaume05} where they showed that this strategy is more efficient than random failures. The networks used to show these results by \cite{guillaume05} had an edge-node ratio of less than 3 and where the networks we use here have a much higher edge-node ratio with the exception of the author network, which has an edge-node ratio of 2.6 and thus we can see differences in the results of random failures and almost random failures in the author network.

\textbf{Targeted attacks on Edges:} All the networks show an equivalent resilience against targeted attack on edges when compared to random removal of edges. The author network in Figure \ref{fig::author}(left) again shows an early breakdown of the biggest component further proving our claim of high mean connectivity being an important reason for resilient structures.

\textbf{Random failure of Edges:} A slight variation in the resilience can be observed for all the networks. All real networks show a tendency to disintegrate more than the generated networks specially after the removal of $60\%$ edges. Author network is the least resilient case where all the generated and the real networks disintegrate into smaller components after a removal of around $60\%$ edges. Since the author network has the least edge-node ratio (see Table \ref{tbl::metrics}), this behavior further proves that other networks show a resilient behavior because of high mean connectivity of the nodes.

\textbf{Almost Random failure of Edges:} All the dataset behave exactly the same except for the case of epinions network where after the removal of $60\%$ edges result in different pattern. The small world and the random network behave similarly as they are least resilient. The scale free and small world-scale free networks behave similarly being more resilient and epinions network is in between these two behaviors.

\begin{figure}
\begin{center}
\includegraphics[width=0.7\textwidth]{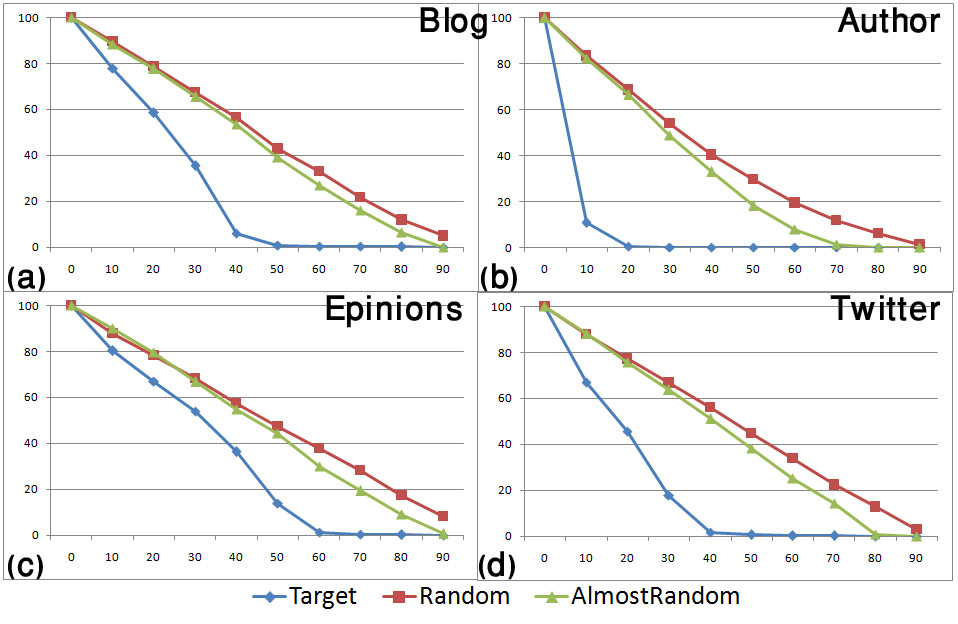}
\end{center}
\vspace{-18pt}
\caption{Comparative analysis of different attack strategies on nodes for the 4 semantically different social networks.}
\label{fig::attackresults}
\end{figure}

The second experiment compares different attack strategies on nodes using four real networks as shown in figure \ref{fig::attackresults}. The previous experiment revealed that targeted attack on nodes is the most efficient attack strategy in terms of different classes of networks. The second experiment compares different attack strategies on nodes for different social networks.

The first findings from this experiment are that there are only minor differences in random attacks and almost random attacks when the edge node ratio of the networks is high. Slight differences can be observed for the author network in figure \ref{fig::attackresults}(b) which has comparatively low edge-node ratio. This is in contradiction to the results of \cite{guillaume05}, who studied internet graphs with much less edge-node ratio. Internet graphs are known to have star-like structures where a single node sits (known as hub) in between many other nodes providing efficient connectivity among many nodes. Removing nodes with degree 2 or more unintentionally targets these hubs which in turn results in breakdown of the network. In contrast to this, social networks do not have hubs. Removing nodes with degree 2 or more does not break the network specially for networks with high edge-node ratio because there are many paths that connect a single node, thus making it more resilient to this type of attack.

The second finding is as expected, the effectiveness of targeted attack on nodes as compared to random and almost random failures. The author networks has a low percolation threshold and the network breaks immediately into relatively larger size connected components. The Blog, Epinions and Twitter network demonstrate a more graceful degradations with a high percolation threshold as most of the nodes remain connected into a single connected component even after the removal of $40\%$ to $60\%$ high degree nodes.

\section{Conclusion}\label{sec::conclusion}

In this paper, we have studied the behavior of small world, scale free and small world-scale free networks in comparison to random and four semantically different social networks. Our results show that that behavior of all these classes of networks remains the same under targeted attack on edges, random attack on nodes and edges, almost random attack on nodes and edges both in terms of size of biggest component and average path length. The behavior of these networks change under targeted attack on nodes. Interesting behavoir was observed on the basis of clustering coefficient and targeted attack on edges. Furthermore structural differences were observed between real social networks and all network generation models. Insignificant differences were observed between random failure of nodes and edges when compared with almost random failures.

We intend to extend this study by incorporating large size social networks. The networks studied are unweighted and undirected, and we intend to analyze the behavior of these networks as well. Another important characteristic of social networks is the temporal dimension which plays an important role in dictating many social processes such as information diffusion and epidemics and we would also like to study resilience for temporal social networks.

\bibliographystyle{abbrv}
\bibliography{visu}

\end{document}